\documentclass[12pt]{article}%
\usepackage{amsmath,latexsym}
\usepackage{graphicx}
\usepackage{amsmath}
\usepackage{amsfonts}
\usepackage{amssymb}
\begin{document}

\title{{Noncommutative-geometry wormholes without
   exotic matter}}
   \author{
Peter K.F. Kuhfittig*\\  \footnote{kuhfitti@msoe.edu}
 \small Department of Mathematics, Milwaukee School of
Engineering,\\
\small Milwaukee, Wisconsin 53202-3109, USA}

\date{}
 \maketitle

\begin{abstract}\noindent
A fundamental property of Morris-Thorne
wormholes is the so-called flare-out
condition that automatically results in
a violation of the null energy condition
(NEC) in classical general relativity.
By contrast, in $f(R)$ modified gravity,
the material threading the wormhole may
actually satisfy the NEC, while, at the
same time, the effective stress-energy \
tensor arising from the modified theory
violates the NEC, thereby sustaining the
wormhole.  It is shown in this paper
that noncommutative geometry, an offshoot
of string theory, can be viewed as a
special case of $f(R)$ gravity and can
therefore sustain a wormhole without
the use of exotic matter.  It also
provides a motivation for the choice
of $f(R)$ in the modified gravitational
theory.\\
\\
\textbf{PAC numbers:}\,\,04.20.Jb, 04.20.-q, 04.50.Kd
\\

\end{abstract}

\section{Introduction}\label{S:introduction}

Wormholes are tunnel-like structures in
spacetime that link widely separated
regions of our Universe or different universes
altogether \cite{MT88}.  This spacetime
geometry can be described by the metric
\begin{equation}\label{E:line}
ds^{2}=-e^{2\Phi(r)}dt^{2}+\frac{dr^2}{1-b(r)/r}
+r^{2}(d\theta^{2}+\text{sin}^{2}\theta\,
d\phi^{2}),
\end{equation}
using units in which $c=G=1$.  In this line
element, $b=b(r)$ is called the \emph{shape
function} and $\Phi=\Phi(r)$ is called the
\emph{redshift function}; the latter must be
finite everywhere to avoid the presence of
an event horizon.  For the shape function we
must have $b(r_0)=r_0$, where $r=r_0$ is the
radius of the \emph{throat} of the wormhole.
Another important requirement is the
\emph{flare-out condition} at the throat:
$b'(r_0)<1$; also, $b(r)<r$ near the throat.
In classical general relativity, the flare-out
condition can only be satisfied by violating
the null energy condition (NEC):
\begin{equation}
  T_{\mu\nu}k^{\mu}k^{\nu}\ge 0
\end{equation}
for all null vectors $k^{\mu}$, where
$T_{\mu\nu}$ is the stress-energy tensor.
In particular, for the outgoing null vector
$(1,1,0,0)$, the violation becomes
\begin{equation}
    T_{\mu\nu}k^{\mu}k^{\nu}=\rho +p_r<0.
\end{equation}
Here $T^t_{\phantom{tt}t}=-\rho$ is the energy
density, $T^r_{\phantom{rr}r}= p_r$ is the
radial pressure, and
$T^\theta_{\phantom{\theta\theta}\theta}=
T^\phi_{\phantom{\phi\phi}\phi}=p_t$ is
the lateral pressure.

According to Lobo \cite{HLMS}, these ideas can
be extended to$f(R)$  modified gravity by
referring back to the Raychaudhury equation.
In this theory, the Ricci scalar $R$ in the
Einstein-Hilbert action
\begin{equation*}
  S_{\text{EH}}=\int\sqrt{-g}\,R\,d^4x
\end{equation*}
is replaced by a nonlinear function $f(R)$:
\begin{equation*}
   S_{f(R)}=\int\sqrt{-g}\,f(R)\,d^4x.
\end{equation*}
To extend these ideas, the stress-energy tensor
$T_{\mu\nu}$ has to be replaced by
$T^{\text{eff}}_{\mu\nu}$, the
\emph{effective} stress-energy tensor
arising from the modified theory,
leading to the Einstein field equations
$G_{\mu\nu}=\kappa^2T^{\text{eff}}_{\mu\nu}$.
The NEC now becomes
\begin{equation}
   T^{\text{eff}}_{\mu\nu}k^{\mu}k^{\nu}\ge 0.
\end{equation}
As a result, the violation of the (generalized)
NEC becomes
$T^{\text{eff}}_{\mu\nu}k^{\mu}k^{\nu}<0$,
which reduces to $T_{\mu\nu}k^{\mu}k^{\nu}<0$
in classical general relativity.  According
to Ref. \cite{HLMS}, it now becomes possible
in principle to allow the matter threading the
wormhole to satisfy the NEC while retaining
the violation of the generalized NEC, i.e.,
$T^{\text{eff}}_{\mu\nu}k^{\mu}k^{\nu}<0$.
So the necessary condition for maintaining
a traversable wormhole has been met.
According to Ref. \cite{HLMS}, the
higher-order curvature terms leading to
the violation may be interpreted as a
gravitational fluid that supports the
wormhole.

The purpose of this paper is to show that
noncommutative geometry, an offshoot of
string theory, is not only an example of
such a modified gravitational theory, it
provides a motivation for the choice of
the function $f(R)$.

\section{Noncommutative geometry}

An important outcome of string theory is
the realization that coordinates may become
noncommutative operators on a $D$-brane
\cite{eW96, SW99}.  Noncommutativity replaces
point-like objects by smeared objects
\cite{SS03, NSS06, NS10} with the aim of
eliminating the divergences that normally
occur in general relativity.  Moreover,
noncommutative geometry results in a
fundamental discretization of spacetime
due to the commutator
$[\textbf{x}^{\mu},\textbf{x}^{\nu}]
=i\theta^{\mu\nu}$, where $\theta^{\mu\nu}$ is
an antisymmetric matrix.

An effective way to model the smearing is
to assume that the energy density of a
static, spherically symmetric, and
particle-like gravitational source has
the form \cite{NM08, LL12}
\begin{equation}\label{E:rho}
  \rho(r)=\frac{\mu\sqrt{\beta}}
     {\pi^2(r^2+\beta)^2}.
\end{equation}
Here the mass $\mu$ is diffused throughout the
region of linear dimension $\sqrt{\beta}$ due
to the uncertainty.  Noncommutative
geometry is an intrinsic property of spacetime
and does not depend on any particular features
such as curvature.  Eq. (\ref{E:rho})
immediately yields the mass distribution
\begin{equation}\label{E:mass}
   m_{\beta}(r)=\int^r_04\pi (r')^2\rho(r')
   dr'=\frac{2M}{\pi}\left(\text{tan}^{-1}
   \frac{r}{\sqrt{\beta}}-
   \frac{r\sqrt{\beta}}{r^2+\beta}\right),
\end{equation}
where $M$ is now the total mass of the source.

\section{Wormholes in modified gravity}

In this section we adopt the point of view that
noncommutative geometry is a modified gravity
theory, but first we make the important
observation that the Einstein field equations
$G_{\mu\nu}=\kappa^2T^{\text{eff}}_{\mu\nu}$
mentioned in Sec. \ref{S:introduction} show
that the noncommutative effects can be
implemented by modifying only the
stress-energy tensor, while leaving the
Einstein tensor unchanged.  As a result,
the length scales can be macroscopic.

The next step is to show that our
noncommutative-geometry background is a
special case of $f(R)$ modified gravity.
To that end, we need to list the gravitational
field equations in the form used by Lobo and
Oliveira \cite{LO09}.  Here we assume that
$\Phi'(r)\equiv 0$; otherwise, according to
Ref. \cite{LO09}, the analysis becomes
intractable.  (It is also assumed that for
notational convenience, $\kappa =1$ in
the field equations.)
\begin{equation}\label{E:Lobo1}
   \rho(r)=F(r)\frac{b'(r)}{r^2},
\end{equation}
\begin{equation}\label{E:Lobo2}
   p_r(r)=-F(r)\frac{b(r)}{r^3}
   +F'(r)\frac{rb'(r)-b(r)}{2r^2}
   -F''(r)\left(1-\frac{b(r)}{r}\right),
\end{equation}
and
\begin{equation}\label{E:Lobo3}
   p_t(r)=-\frac{F'(r)}{r}\left(1-\frac{b(r)}{r}
   \right)+\frac{F(r)}{2r^3}[b(r)-rb'(r)],
\end{equation}
where $F=\frac{df}{dR}$.  If $F(r)\equiv 1$,
then Eqs. (\ref{E:Lobo1}) - (\ref{E:Lobo3})
reduce to the usual field equations with
$\kappa =1$ and $\Phi'(r)\equiv 0$.
So from $\rho(r)=b'(r)/r^2$ and Eq.
(\ref{E:rho}), we obtain the shape function
\begin{equation}\label{E:shape}
  b(r)=\frac{M\sqrt{\beta}}{2\pi^2}
  \left(\frac{1}{\sqrt{\beta}}\text{tan}^{-1}
  \frac{r}{\sqrt{\beta}}-\frac{r}{r^2+\beta}-
  \frac{1}{\sqrt{\beta}}\text{tan}^{-1}
  \frac{r_0}{\sqrt{\beta}}+\frac{r_0}{r_0^2
  +\beta}\right)+r_0,
\end{equation}
where $M$ is now the mass of the wormhole,
and from
\begin{equation}\label{E:bprime}
   b'(r)=\frac{M\sqrt{\beta}}{\pi^2}
     \frac{r^2}{(r^2+\beta)^2},
\end{equation}
we see that $b'(r_0)<1$; the flare-out
condition is thereby met.

According to Ref. \cite{LO09}, the Ricci
scalar is
\begin{equation}\label{E:R1}
   R(r)=\frac{2b'(r)}{r^2}.
\end{equation}


\section{Avoiding exotic matter}

Since we wish the matter threading the
wormhole to obey the null energy
condition, we require that $\rho + p_r
\ge 0$ and $\rho +p_t\ge 0$, as well
as $\rho\ge 0$.  From Eqs. (\ref{E:Lobo1})
and (\ref{E:Lobo2}), we therefore need to
satisfy the following conditions:
\begin{equation}\label{E:Con1}
   \rho=\frac{Fb'}{r^2}\ge 0
\end{equation}
and
\begin{equation}\label{E:Con2}
    \rho +p_r=
   \frac{(2F+rF')(b'r-b)}{2r^3}
   -F''\left(1-\frac{b}{r}\right)
   \ge 0.
\end{equation}
Using Eqs. (\ref{E:Lobo1}) and
(\ref{E:rho}), we get
\begin{equation}\label{E:F(r)}
   F(r)=\frac{r^2}{b'(r)}\rho(r)=
   \frac{1}{b'(r)/r^2}
   \frac{\mu\sqrt{\beta}}{\pi^2}
   \frac{1}{(r^2+\beta)^2}.
\end{equation}
Eq. (\ref{E:R1}) now implies that
\begin{equation}\label{E:r}
   r(R)=\sqrt{\frac{2b'}{R}}.
\end{equation}
Substituting in Eq. (\ref{E:F(r)}) yields
\begin{equation}\label{E:F(R)}
  F(R)=\frac{2\mu\sqrt{\beta}}{\pi^2}
  \frac{1}{R\left(\frac{2b'}{R}+\beta
  \right)^2}
\end{equation}
and
\begin{equation}\label{E:derivative}
   F'(R)=-\frac{4\mu\sqrt{\beta}}{\pi^2}
   \frac{2b'+2\beta R}{(2b'R+\beta R^2)^3}.
\end{equation}
Inequality (\ref{E:Con1}) is evidently
satisfied.  Substituting in Inequality
(\ref{E:Con2}), we obtain (at or near the
throat)
\begin{equation}\label{E:Omega1}
   \rho+p_r=\frac{(2F+rF')(b'r-b)}{2r^3}
   =\frac{1}{r^3}\frac{2\mu\sqrt{\beta}}{\pi^2}
   \frac{1}{(2b'R+\beta R^2)^2}\left(1-
   \frac{r}{R}\frac{2b'+2\beta R}{2b'+\beta R}
   \right)(b'r-b).
\end{equation}
To show that $\rho+p_r|_{r=r_0} > 0$,
recall from Eq. (\ref{E:bprime}) that
$b'(r_0)\ll 1$ since $\beta$ is
extremely small.  So
\begin{equation}
   \left.\frac{r}{R}\right |_{r=r_0}=
    \frac{r_0^3}{2b'(r_0)}>1,
\end{equation}
as long as the throat size is not
microscopic, i.e., $r_0>[2b'(r_0)]^{1/3}$.
Since we also have
$(2b'+2\beta R)(2b'+\beta R)>1$, it now follows
that
\begin{equation}\label{E:Omega2}
   \rho+p_r|_{r=r_0} > 0.
\end{equation}
Next, we find that
\begin{equation}
    \rho+p_t|_{r=r_0}=F(r_0)\frac{b'(r_0)+1}
       {2r_0^2}>0.
\end{equation}
The NEC is therefore satisfied at the throat
for the null vectors $(1,1,0,0)$,
$(1,0,1,0)$, and $(1,0,0,1)$.  It is
shown in Ref. \cite{pK18} that the result
can be extended to any null vector
\[
   (1,a,b,c),\quad 0\le a,b,c\le 1,\quad
   a^2+b^2+c^2=1.
\]

Since $F=\frac{df}{dR}$, Eq. (\ref{E:F(R)})
also yields
\begin{equation}\label{E:f(R)}
    f(R)=\int^R_0\frac{2\mu\sqrt{\beta}}{\pi^2}
    \frac{1}{R'\left(\frac{2b'}{R'}+\beta
    \right)^2}dR'=
    \frac{2\mu\sqrt{\beta}}{\pi^2}
    \frac{(\beta R+2b')\text{ln}\,
    (\beta R+2b')-\beta R}
    {\beta^2(\beta R+2b')}+C.
\end{equation}

To check the violation of the generalized
NEC, i.e.,
$T^{\text{eff}}_{\mu\nu}k^{\mu}k^{\nu}<0$,
we follow Lobo and Oliveira \cite {LO09}:
\begin{multline}
  \left. \rho^{\text{eff}}+p_r^{\text{eff}}
  |_{r=r_0}
   =\frac{1}{F}\frac{rb'-b}{r^3}+
   \frac{1}{F}\left(1-\frac{b}{r}\right)
   \left(F''-F'\frac{b'r-b}{2r^2(1-b/r}
   \right)\right|_{r-r_0}\\
   =\frac{1}{F}\frac{b'(r_0)-1}
   {r_0^2}+\frac{1-b'(r_0)}{2r_0}
   \frac{F'}{F}.
\end{multline}
Since $F'(R)<0$, it follows that
\begin{equation}
  \rho^{\text{eff}}+p_r^{\text{eff}}
   |_{r=r_0}<0.
\end{equation}
So the generalized NEC is violated
thanks to the stress-energy tensor
$T^{\text{eff}}_{\mu\nu}$.

\section{The high radial tension}\label{S:tension}

Although not part of this study, noncommutative
geometry plays another important role in
wormhole physics.  According to Ref. \cite{MT88},
for a moderately-sized wormhole, the radial
tension at the throat has the same magnitude
as the pressure at the center of a massive
neutron star.  Attributing this outcome to
exotic matter is rather problematical since
exotic matter was introduced primarily to
ensure the violation of the null energy
condition.

It is shown in Ref. \cite{pK20} that such an
outcome can be accounted for by the
noncommutative geometry background.
Recalling that noncommutative geometry is
an offshoot of string theory, this approach
can be viewed as a foray into quantum
gravity.

\section{Conclusion}

A fundamental geometric property of traversable
wormholes is the flare-out condition $b'(r_0)<1$.
In classical general relativity, the flare-out
condition can only be met by violating the NEC,
$T_{\mu\nu}k^{\mu}k^{\nu}<0$, for all null
vectors $k^{\mu}$.  In $f(R)$ modified gravity,
the stress-energy tensor $T_{\mu\nu}$ is
replaced by the effective stress-energy tensor
$T^{\text{eff}}_{\mu\nu}$ arising from the
modified theory.  So it is possible in principle
to have a violation of the generalized NEC,
$ T^{\text{eff}}_{\mu\nu}k^{\mu}k^{\nu}<0$,
while maintaining the NEC,
$T_{\mu\nu}k^{\mu}k^{\nu}\ge 0$, for the
material threading the wormhole.  According
to Ref. \cite{HLMS}, the higher-order
curvature terms leading to the violation
may be interpreted as a gravitational fluid
that supports the wormhole.

The purpose of this paper is to show that
noncommutative geometry, an offshoot of
string theory, can be viewed as a special
case of $f(R)$ modified gravity, where
$f(R)$ is given by Eq. (\ref{E:f(R)}).
The result is a zero-tidal force
traversable wormhole without exotic matter.
The noncommutative-geometry background
also provides a motivation for the choice
of $f(R)$.

Sec. \ref{S:tension} reiterates another
aspect of noncommutative geometry, the
ability to account for the enormous
radial tension in a Morris-Thorne
wormhole, as shown in Ref. \cite{pK20}.


\begin{thebibliography}{20}

\bibitem{MT88}M.S. Morris and K.S. Thorne, Wormholes in
   spacetime and their use for interstellar travel: A tool
   for teaching general relativity, Amer. J. Phys.,
   56 (1988) 395-412.
\bibitem{HLMS}T. Harko, F.S.N. Lobo, M.K. Mak, and S.V.
   Sushkov, Modified-gravity wormholes without exotic
   matter, Phys. Rev. D, 87 (2013) 067504.
\bibitem{eW96}E. Witten, Bound states of strings and
   $p$-branes, Nucl. Phys. B, 460 (1996) 335-350.
\bibitem{SW99}N. Seiberg and E. Witten, String theory
   and noncommutative geometry, J. High Energy Phys.,
    9909 (1999) 032.
\bibitem{SS03}A. Smailagic and E. Spallucci, Feynman
   path integral on the non-commutative plane, J. Phys. A,
   36 (2003) L-467-L-471.
\bibitem{NSS06}P. Nicolini, A. Smailagic, and E. Spallucci,
   Noncommutative geometry inspired Schwarzschild black
   hole, Phys. Lett. B, 632 (2006) 547-551.
\bibitem{NS10}P. Nicolini and E. Spallucci,
   Noncommutative geometry-inspired dirty black holes,
   Class. Quant. Grav., 27 (2010) 015010.
\bibitem{NM08}K. Nozari and S.H. Mehdipour, Hawking
   radiation as quantum tunneling for a noncommutative
   Schwarzschild black hole, Class. Quant. Grav.,
   25 (2008) 175015.
\bibitem{LL12}J. Liang and B. Liu, Thermodynamics of
   noncommutative geometry inspired BTZ black hole based
   on Lorentzian smeared mass distribution, Europhys.
   Lett., 100 (2012) 30001.
\bibitem{LO09}F.S.N. Lobo and M.A. Oliveira, Wormhole
   geometries in $f(R)$ modified theories of gravity,
   Phys. Rev. D, 80 (2009) 104012.
\bibitem{pK18}P.K.F. Kuhfittig, ``Traversable wormholes
   sustained by an extra spatial dimension,"
   Phys. Rev. D, 98 (2018) 064041.
\bibitem{pK20}P.K.F. Kuhfittig, Accounting for the large
   radial tension in Morris-Thorne wormholes,
   Eur. Phys. J. Plus, 135 (2020) 50.

 \end{thebibliography}
\end{document}